\documentstyle[aps,preprint]{revtex}

\begin{document}
\author{Remo Garattini}
\address{M\'ecanique et Gravitation, Universit\'e de Mons-Hainaut,\\
Facult\'e des Sciences, 15 Avenue Maistriau, \\
B-7000 Mons, Belgium \\
and\\
Facolt\`a di Ingegneria, Universit\`a degli Studi di Bergamo,\\
Viale Marconi, 5, 24044 Dalmine (Bergamo) Italy\\
e-mail: Garattini@mi.infn.it}
\title{Probing foamy spacetime with Variational Methods}
\date{\today}
\maketitle

\begin{abstract}
A one-loop correction of the quasilocal energy in the Schwarzschild
background, with flat space as a reference metric, is performed by means of
a variational procedure in the Hamiltonian framework. We examine the
graviton sector in momentum space, in the lowest possible state. An
application to the black hole pair creation via the Casimir energy is
presented. Implications on the foam-like scenario are discussed.
\end{abstract}

\section{Introduction}

It is generally accepted that if Einstein gravity has a ground state, this
is represented by Flat space. Many attempts to discover a possible lower
minimum different from the Flat one have been carried out and it seems
nowadays that, at least at classical level, the lowest energy configuration
state with $E=0$ is attributed to Flat space\cite{Yau}. From the quantum
mechanical point of view we can obtain information on the stability of the
spacetime by means of saddle point methods: if a lower state exists, this
can be reached by quantum tunneling. As a consequence a negative mode will
appear in the Lichnerowicz operator. Actually a tunneling process was
discovered by {\it Gross, Perry }and {\it Yaffe}\cite{GPY}, but it involves
a flat space having a temperature $T\neq 0$, which means that we have not
considered as initial state the correct ``{\it vacuum}''. Thus also this
result seems to corroborate the stability of flat space under quantum
fluctuation via nucleation of black holes. Recently a different mechanism
involving both a topology change and a black hole pair production has been
considered\cite{Remo}. In this paper we are investigating the possibility of
computing such effects in a Hamiltonian approach in presence of quasilocal
energy. To this purpose we begin by fixing the background manifold ${\cal M}$
which is represented by the Schwarzschild spacetime. In particular we are
interested in a constant time section of ${\cal M}$, best known as
Einstein-Rosen bridge with wormhole topology $R^1\times S^2$, defining a
bifurcation surface which divides $\Sigma $ into two parts denoted by $%
\Sigma _{+}$ and $\Sigma _{-}$. Our purpose is to consider perturbations at $%
\Sigma $, which naturally define quantum fluctuations of the Einstein-Rosen
bridge and evaluate quasilocal energy corrections to one-loop with the
assumption of neglecting quantum excitations on the boundaries, motivated by
the fact that in asymptotic spacelike directions, quasilocal energy
approaches the ${\cal ADM}$ term defined in the asymptotic region, whose
quantum fluctuations are unphysical. Nevertheless this approximation is not
valid when we consider finite distance located boundaries and
multi-wormholes configuration\cite{BrownYork,FroMar}. In this context, we
will examine the effect of perturbations on the possible ``{\it ground state}%
'' by means of a variational approach applied on gaussian wave functional.
Indeed, if we discover the existence of negative modes in this
approximation, it is quite reasonable to think of a tunneling process which
moves spacetime from the false vacuum to the true one. The rest of the paper
is structured as follows, in section \ref{p1}, we have borrowed from Refs.%
\cite{FroMar,HawHor} the general expressions for Hamiltonian and quasilocal
energy, in section \ref{p2}, we analyze the stationary Schr\"{o}dinger
equation coming from the perturbed wormhole metric and we give some of the
basic rules to perform the functional integration for the Hamiltonian
approximated to second order, in section \ref{p3}, we analyze the spin-2
operator or the operator acting on transverse traceless tensors. We
summarize and conclude in section \ref{p4}.

\section{Quasilocal Energy and Energy Density Calculation in Schr\"{o}dinger
Representation}

\label{p1}Although it is not necessary for the forthcoming discussions, let
us consider the maximal analytic extension of the Schwarzschild metric,
i.e., the Kruskal manifold whose spatial slices $\Sigma $ represent
Einstein-Rosen bridges with wormhole topology $S^2\times R^1$. Following Ref.%
\cite{FroMar}, the complete manifold ${\cal M}$ can be taken as a model for
an eternal black hole composed of two wedges ${\cal M}_{+}$ and ${\cal M}_{-}
$ located in the right and left sectors of a Kruskal diagram. The
hypersurface $\Sigma $ is divided in two parts $\Sigma _{+}$ and $\Sigma _{-}
$ by a bifurcation two-surface $S_0$. On $\Sigma $ we can write the
gravitational Hamiltonian 
\[
H_p=H-H_0=\int_\Sigma d^3x(N{\cal H+}N^i{\cal H}_i)
\]
\begin{equation}
+\text{ }\frac 1\kappa \int_{S_{+}}^{}d^2xN\sqrt{\sigma }\left( k-k^0\right)
-\frac 1\kappa \int_{S_{-}}d^2xN\sqrt{\sigma }\left( k-k^0\right) ,
\label{a1}
\end{equation}
where $\kappa =8\pi G$. The Hamiltonian has both volume and boundary
contributions. The volume part involves the Hamiltonian and momentum
constraints 
\[
{\cal H}=\left( 2\kappa \right) G_{ijkl}\frac{\pi ^{ij}\pi ^{kl}}{\sqrt{^3g}}%
-\sqrt{^3g}R/\left( 2\kappa \right) =0,
\]
\begin{equation}
{\cal H}_i=-2\pi _{i|j}^j=0,
\end{equation}
where $G_{ijkl}=\frac 12\left( g_{ik}g_{jl}+g_{il}g_{jk}-g_{ij}g_{kl}\right) 
$ is the supermetric and $R$ denotes the scalar curvature of the surface $%
\Sigma $. The volume part of the Hamiltonian $\left( \ref{a1}\right) $ is
zero when the Hamiltonian and momentum constraints are imposed. However, for
the flat and the Schwarzschild space, constraints are immediately satisfied,
then in this context the total Hamiltonian reduces to 
\begin{equation}
H_p=\frac 1\kappa \int_{S_{+}}^{}d^2xN\sqrt{\sigma }\left( k-k^0\right) -%
\frac 1\kappa \int_{S_{-}}d^2xN\sqrt{\sigma }\left( k-k^0\right) .
\label{a1a}
\end{equation}
Quasilocal energy is defined as the value of the Hamiltonian that generates
unit time translations orthogonal to the two-dimensional boundaries, i.e. 
\[
E_{{\rm tot}}=H_{{\rm quasilocal}}=H_{+}-H_{-}=E_{+}-E_{-},
\]
\[
E_{+}=\frac 1\kappa \int_{S_{+}}^{}d^2x\sqrt{\sigma }\left( k-k^0\right) 
\]
\begin{equation}
E_{-}=-\frac 1\kappa \int_{S_{-}}d^2x\sqrt{\sigma }\left( k-k^0\right) .
\end{equation}
where $\left| N\right| =1$ at both $S_{+}$ and $S_{-}$. $E_{tot}$ is the
quasilocal energy of a spacelike hypersurface $\Sigma =\Sigma _{+}\cup
\Sigma _{-}$ bounded by two boundaries $^3S_{+}$ and $^3S_{-}$ located in
the two disconnected regions $M_{+}$ and $M_{-}$ respectively. We have
included the subtraction terms $k^0$ for the energy. $k^0$ represents the
trace of the extrinsic curvature corresponding to embedding in the
two-dimensional boundaries $^2S_{+}$ and $^2S_{-}$ in three-dimensional
Euclidean space. Following the same scheme of the boundary subtraction
procedure, we would like to discuss the possibility of generalizing such a
procedure. To this end, by looking at the Hamiltonian structure, we see that
there are two classical constraints 
\begin{equation}
\left\{ 
\begin{array}{l}
{\cal H}\text{ }=0 \\ 
{\cal H}^i=0
\end{array}
\right. ,
\end{equation}
which are satisfied both by the Schwarzschild and Flat metric and two {\it %
quantum} constraints 
\begin{equation}
\left\{ 
\begin{array}{l}
{\cal H}\tilde{\Psi}\text{ }=0 \\ 
{\cal H}^i\tilde{\Psi}=0
\end{array}
\right. .
\end{equation}
${\cal H}\tilde{\Psi}$ $=0$ is known as the {\it Wheeler-DeWitt} equation
(WDW). Nevertheless, we are interested in assigning a meaning to 
\begin{equation}
\frac{\left\langle \Psi \left| H^{{\rm Schw}.}-H^{{\rm Flat}}\right| \Psi
\right\rangle }{\left\langle \Psi |\Psi \right\rangle }+\frac{\left\langle
\Psi \left| H_{{\rm quasilocal}}\right| \Psi \right\rangle }{\left\langle
\Psi |\Psi \right\rangle },  \label{a2}
\end{equation}
where $\Psi $ is a wave functional whose structure will be determined later
and $H^{{\rm Schw.}}\left( H^{{\rm Flat}}\right) $ is the total Hamiltonian
referred to the different spacetimes. Note that the first term of $\left( 
\ref{a1}\right) $ is simply the extension of the quasilocal energy
subtraction procedure generalized to the volume term. Note also that if the
expectation value is calculated on the wave functional solution of the WDW
equation, we obtain only the boundary contribution. Then to give meaning to $%
\left( \ref{a1}\right) $, we adopt the semiclassical strategy of the WKB
expansion. By observing that the kinetic part of the Super Hamiltonian is
quadratic in the momenta, we expand the three-scalar curvature $\int d^3x%
\sqrt{g}R^{\left( 3\right) }$ up to quadratic order and we get 
\begin{equation}
\int d_{}^3x\left[ -\frac 14h\triangle h+\frac 14h^{li}\triangle h_{li}-%
\frac 12h^{ij}\nabla _l\nabla _ih_j^l+\frac 12h\nabla _l\nabla _ih_{}^{li}-%
\frac 12h^{ij}R_{ia}h_j^a+\frac 12hR_{ij}h_{}^{ij}\right] ,
\end{equation}
where $h$ is the trace of $h_{ij}$. On the other hand, following the usual
WKB expansion, we will consider $\tilde{\Psi}\simeq C\exp \left( iS\right) $%
. In this context, the approximated wave functional will be substituted by a 
{\it trial wave functional} according to the variational approach we would
like to implement as regards this problem.

\section{Gaussian Wave Functional and Energy Density Calculation in
Schr\"{o}dinger Representation}

\label{p2}

To actually make such calculations, we need an orthogonal decomposition for
both $\pi _{ij\text{ }}^{}$and $h_{ij}^{}$ to disentangle gauge modes from
physical deformations. We define the inner product

\begin{equation}
\left\langle h,k\right\rangle :=\int_\Sigma \sqrt{g}G^{ijkl}h_{ij}\left(
x\right) k_{kl}\left( x\right) d^3x,
\end{equation}
by means of the inverse WDW metric $G_{ijkl}$, to have a metric on the space
of deformations, i.e. a quadratic form on the tangent space at h, with

\begin{equation}
\begin{array}{c}
G^{ijkl}=(g^{ik}g^{jl}+g^{il}g^{jk}-2g^{ij}g^{kl})\text{.}
\end{array}
\end{equation}
The inverse metric is defined on co-tangent space and it assumes the form

\begin{equation}
\left\langle p,q\right\rangle :=\int_\Sigma \sqrt{g}G_{ijkl}p^{ij}\left(
x\right) q^{kl}\left( x\right) d^3x\text{,}
\end{equation}
so that

\begin{equation}
G^{ijnm}G_{nmkl}=\frac 12\left( \delta _k^i\delta _l^j+\delta _l^i\delta
_k^j\right) .
\end{equation}
Note that in this scheme the ``inverse metric'' is actually the WDW metric
defined on phase space. Now, we have the desired decomposition on the
tangent space of 3-metric deformations\cite{BerEbi,York}:

\begin{equation}
h_{ij}=\frac 13hg_{ij}+\left( L\xi \right) _{ij}+h_{ij}^{\bot }  \label{b0}
\end{equation}
where the operator $L$ maps $\xi _i$ into symmetric tracefree tensors

\begin{equation}
\left( L\xi \right) _{ij}=\nabla _i\xi _j+\nabla _j\xi _i-\frac 23%
g_{ij}\left( \nabla \cdot \xi \right) .
\end{equation}
Then the inner product between three-geometries becomes 
\[
\left\langle h,h\right\rangle :=\int_\Sigma \sqrt{g}G^{ijkl}h_{ij}\left(
x\right) h_{kl}\left( x\right) d^3x=
\]
\begin{equation}
\int_\Sigma \sqrt{g}\left[ -\frac 23h^2+\left( L\xi \right) ^{ij}\left( L\xi
\right) _{ij}+h^{ij\bot }h_{ij}^{\bot }\right] .  \label{b1}
\end{equation}
With the orthogonal decomposition in hand we can define a ``{\it Vacuum
Trial State}'' 
\begin{equation}
\Psi \left[ h_{ij}\left( \overrightarrow{x}\right) \right] ={\cal N}\exp
\left\{ -\frac 1{4l_p^2}\left[ \left\langle hK^{-1}h\right\rangle
_{x,y}^{\bot }+\left\langle \left( L\xi \right) K^{-1}\left( L\xi \right)
\right\rangle _{x,y}^{\Vert }+\left\langle hK^{-1}h\right\rangle
_{x,y}^{Trace}\right] \right\} ,
\end{equation}
which will be used as a probe for the gravitational ground state. This
particular expression is useful because the functional can be represented as
a product of three functionals defined on the decomposed tensor field 
\begin{equation}
\Psi \left[ h_{ij}\left( \overrightarrow{x}\right) \right] ={\cal N}\Psi
\left[ h_{ij}^{\bot }\left( \overrightarrow{x}\right) \right] \Psi \left[
\left( L\xi \right) _{ij}\right] \Psi \left[ \frac 13g_{ij}h\left( 
\overrightarrow{x}\right) \right] .  \label{b1a}
\end{equation}
$h_{ij}^{\bot }$ is the tracefree-transverse part of the $3D$ quantum field, 
$\left( L\xi \right) _{ij}$ is the longitudinal part and finally $h$ is the
trace part of the same field. $\left\langle \cdot ,\cdot \right\rangle _{x,y}
$ denotes space integration and $K^{-1}$ is the inverse propagator
containing variational parameters. The main reason for a similar ``{\it %
Ansatz}'' comes from the observation that the quadratic part in the momenta
of the Hamiltonian decouples in the same way of eq.$\left( \ref{b1}\right) $%
. Note that the decomposition related to the momenta is independent of the
choice of the functional. To calculate the energy density, we need to know
the action of some basic operators on $\Psi \left[ h_{ij}\right] $. The
action of the operator $h_{ij}$ on $|\Psi \rangle =\Psi \left[ h_{ij}\right] 
$ is realized by 
\begin{equation}
h_{ij}\left( x\right) |\Psi \rangle =h_{ij}\left( \overrightarrow{x}\right)
\Psi \left[ h_{ij}\right] .
\end{equation}
The action of the operator $\pi _{ij}$ on $|\Psi \rangle $, in general, is

\begin{equation}
\pi _{ij}\left( x\right) |\Psi \rangle =-i\frac \delta {\delta h_{ij}\left( 
\overrightarrow{x}\right) }\Psi \left[ h_{ij}\right] .
\end{equation}
The inner product is defined by the functional integration: 
\begin{equation}
\left\langle \Psi _1\mid \Psi _2\right\rangle =\int \left[ {\cal D}%
h_{ij}\right] \Psi _1^{*}\left\{ h_{ij}\right\} \Psi _2\left\{
h_{kl}\right\} ,
\end{equation}
and the energy eigenstates satisfy the stationary Schr\"{o}dinger equation: 
\begin{equation}
\int d^3x{\cal H}\left\{ -i\frac \delta {\delta h_{ij}\left( \overrightarrow{%
x}\right) },h_{ij}\left( \overrightarrow{x}\right) \right\} \Psi \left\{
h_{ij}\right\} =E\Psi \left\{ h_{ij}\right\} ,  \label{b2}
\end{equation}
where ${\cal H}\left\{ -i\frac \delta {\delta h_{ij}\left( x\right) }%
,h_{ij}\left( x\right) \right\} $ is the Hamiltonian density. Note that the
previous equation in the general context of Einstein gravity is devoid of
meaning, because of the constraints. However in the semiclassical context,
we can give a meaning to eq.$\left( \ref{b2}\right) $, where a {\it %
semiclassical time} is introduced in the same manner of Refs.\cite
{HalHaw,Halliwell}. There, a Schr\"{o}dinger equation of the form 
\begin{equation}
i\frac{\partial \Psi ^{\bot }}{\partial t}=H_{|2}\Psi ^{\bot }  \label{b3a}
\end{equation}
is recovered by the WDW equation approximated to second order for a
perturbed minisuperspace Friedmann model without boundary terms. When
asymptotically flat boundary terms are present we have to take account of
such contributions in the WKB expansion such as in Ref.\cite{Brotz}. However
in this paper only gravitational transverse-traceless modes are considered
on the fixed curved background and $\Psi ^{\bot }$ is substituted by a trial
wave functional. To further proceed, instead of solving $\left( \ref{b2}%
\right) $, which is of course impossible, we can formulate the same problem
by means of a variational principle. We demand that 
\begin{equation}
\frac{\left\langle \Psi \mid H_{|2}\mid \Psi \right\rangle }{\left\langle
\Psi \mid \Psi \right\rangle }=\frac{\int \left[ {\cal D}g_{ij}^{\bot
}\right] \int d_{}^3x\Psi ^{*}\left\{ g_{ij}^{\bot }\right\} {\cal H}%
_{|2}\Psi \left\{ g_{kl}^{\bot }\right\} }{\int \left[ {\cal D}g_{ij}^{\bot
}\right] \mid \Psi \left\{ g_{ij}^{\bot }\right\} \mid ^2}  \label{b2a}
\end{equation}
be stationary against arbitrary variations of $\Psi \left\{ h_{ij}\right\} $%
. The form of $\left\langle \Psi \mid H_{|2}\mid \Psi \right\rangle $ can be
computed as follows. We define normalized mean values 
\begin{equation}
\bar{g}_{ij}^{\bot }\left( \overrightarrow{x}\right) =\frac{\int \left[ 
{\cal D}g_{ij}^{\bot }\right] \int d_{}^3xg_{ij}^{\bot }\left( 
\overrightarrow{x}\right) \mid \Psi \left\{ g_{ij}^{\bot }\right\} \mid ^2}{%
\int \left[ {\cal D}g_{ij}^{\bot }\right] \mid \Psi \left\{ g_{ij}^{\bot
}\right\} \mid ^2},
\end{equation}
\begin{equation}
\bar{g}_{ij}^{\bot }\left( \overrightarrow{x}\right) \text{ }\bar{g}%
_{kl}^{\bot }\left( \overrightarrow{y}\right) +K_{ijkl^{}}^{\bot }\left( 
\overrightarrow{x},\overrightarrow{y}\right) 
\end{equation}
\begin{equation}
=\frac{\int \left[ {\cal D}g_{ij}^{\bot }\right] \int d_{}^3xg_{ij}^{\bot
}\left( \overrightarrow{x}\right) g_{kl}^{\bot }\left( \overrightarrow{y}%
\right) \mid \Psi \left\{ g_{ij}^{\bot }\right\} \mid ^2}{\int \left[ {\cal D%
}g_{ij}^{\bot }\right] \mid \Psi \left\{ g_{ij}^{\bot }\right\} \mid ^2}.
\end{equation}
It follows that, by defining $h_{ij}^{\bot }=g_{ij}-\bar{g}_{ij}$, we have 
\begin{equation}
\int \left[ {\cal D}h_{ij}^{\bot }\right] h_{ij}^{\bot }\left( 
\overrightarrow{x}\right) \mid \Psi \left\{ h_{ij}^{\bot }+\bar{g}%
_{ij}^{\bot }\right\} \mid ^2=0  \label{b3}
\end{equation}
and 
\[
\int \left[ {\cal D}h_{ij}^{\bot }\right] \int d_{}^3xh_{ij}^{\bot }\left( 
\overrightarrow{x}\right) h_{kl}^{\bot }\left( \overrightarrow{y}\right)
\mid \Psi \left\{ h_{ij}^{\bot }+\bar{g}_{ij}^{\bot }\right\} \mid ^2=
\]
\begin{equation}
K_{ijkl^{}}^{\bot }\left( \overrightarrow{x},\overrightarrow{y}\right) \int
\left[ {\cal D}h_{ij}^{\bot }\right] \mid \Psi \left\{ h_{ij}^{\bot }+\bar{g}%
_{ij}^{\bot }\right\} \mid ^2.  \label{b4}
\end{equation}
Nevertheless, the application of the variational principal on arbitrary wave
functional does not improve the situation described by the eq.$\left( \ref
{b2}\right) $. To this purpose, we give to the trial wave functional the
form 
\begin{equation}
\Psi \left[ h_{ij}^{\bot }\right] ={\cal N}\exp \left\{ -\frac 1{4l_p^2}%
\left\langle \left( g-\overline{g}\right) K^{-1}\left( g-\overline{g}\right)
\right\rangle _{x,y}^{\bot }\right\} .  \label{b4a}
\end{equation}
We immediately conclude that 
\begin{equation}
\left\langle \Psi |\pi _{ij}^{\bot }\left( \overrightarrow{x}\right) |\Psi
\right\rangle =0
\end{equation}
where $\pi _{ij}^{\bot }$ is the TT momentum. In Appendix \ref{p6}, we will
show that 
\begin{equation}
\left\langle \Psi |\pi _{ij}^{\bot }\left( \overrightarrow{x}\right) \pi
_{kl}^{\bot }\left( \overrightarrow{y}\right) |\Psi \right\rangle =\frac 14%
K_{ijkl}^{-1}\left( \overrightarrow{x},\overrightarrow{y}\right) .
\label{b5}
\end{equation}
Choice $\left( \ref{b4a}\right) $ is related to the form of the Hamiltonian
approximated to quadratic order in the metric deformations. Indeed, up to
this order we have a harmonic oscillator whose ground state has a Gaussian
form. By means of decomposition $\left( \ref{b0}\right) $, we extract the TT
sector contribution in the previous expression. Moreover, the functional
representation $\left( \ref{b1a}\right) $ eliminates every interaction
between gauge and the other terms. Then for the TT sector (spin-two), one
gets 
\begin{equation}
\int_\Sigma d^3x\sqrt{g}R^{\left( 3\right) }\simeq \frac 1{4l_p^2}%
\int_\Sigma d^3x\sqrt{g}\left[ h^{\bot ij}\left( \triangle _2\right)
_j^ah_{ia}^{\bot }-2hR_{ij}h^{\bot ij}\right] ,  \label{b5a}
\end{equation}
where $\left( \triangle _2\right) _j^a:=-\triangle \delta _j^a+2R_j^a$. The
latter term disappears because the gaussian integration does not mix the
components. Then by collecting together eq.$\left( \ref{b5a}\right) $ and eq.%
$\left( \ref{b5}\right) $, one obtains the one-loop-like Hamiltonian form
for TT deformations 
\begin{equation}
H^{\bot }=\frac 1{4l_p^2}\int_\Sigma ^{}d^3x\sqrt{g}G^{ijkl}\left[ K^{-1\bot
}\left( x,x\right) _{ijkl}+\left( \triangle _2\right) _j^aK^{\bot }\left(
x,x\right) _{iakl}\right] .  \label{b6}
\end{equation}
The propagator $K^{\bot }\left( x,x\right) _{iakl}$ comes from a functional
integration and it can be represented as 
\begin{equation}
K^{\bot }\left( \overrightarrow{x},\overrightarrow{y}\right) _{iakl}:=\sum_N%
\frac{h_{ia}^{\bot }\left( \overrightarrow{x}\right) h_{kl}^{\bot }\left( 
\overrightarrow{y}\right) }{2\lambda _N\left( p\right) },
\end{equation}
where $h_{ia}^{\bot }\left( \overrightarrow{x}\right) $ are the
eigenfunctions of $\triangle _{2j}^a$ and $\lambda _N\left( p\right) $ are
infinite variational parameters.

\section{The Spectrum of the Spin-2 Operator and the evaluation of the
Energy Density in Momentum Space}

\label{p3}

The Spin-two operator is defined by

\begin{equation}
\left( \triangle _2\right) _j^a:=-\triangle \delta _j^{a_{}^{}}+2R_j^a
\end{equation}
where $\triangle $ is the curved Laplacian (Laplace-Beltrami operator) on a
Schwarzschild background and $R_{j\text{ }}^a$ is the mixed Ricci tensor
whose components are:

\begin{equation}
R_j^a=diag\left\{ \frac{-2m}{r_{}^3},\frac m{r_{}^3},\frac m{r_{}^3}\right\}
,
\end{equation}
where $2m=2MG$. This operator is similar to the Lichnerowicz operator
provided that we substitute the Riemann tensor with the Ricci tensor. This
is essentially due to the fact that the Riemann tensor in three-dimensions
is a linear combination of the Ricci tensor. In $\left( \ref{d1}\right) $
the Ricci tensor acts as a potential on the space of TT tensors; for this
reason we are led to study the following eigenvalue equation

\begin{equation}
\left( -\triangle \delta _j^a+2R_j^a\right) h_a^i=E^2h_j^{i_{}^{}}
\label{d1}
\end{equation}
where $E^2$ is the eigenvalue of the corresponding equation. In doing so, we
follow Regge and Wheeler in analyzing the equation as modes of definite
frequency, angular momentum and parity. We can specialize to the case with $%
M=0$ without altering the contribution to the total energy because of the
spherical symmetry of the problem. We recall that $L$ is the quantum number
corresponding to the square of angular momentum and $M$ is the quantum
number corresponding to the projection of the angular momentum on the
z-axis. In this case, Regge-Wheeler decomposition \cite{Regge} shows that
the even-parity three-dimensional perturbation is

\begin{equation}
h_{ij}^{even}\left( r,\vartheta ,\phi \right) =diag\left[ H\left( r\right)
\left( 1-\frac{2m}r\right) ^{-1},r^2K\left( r\right) ,r^2\sin ^2\vartheta
K\left( r\right) \right] Y_{l0}\left( \vartheta ,\phi \right) .  \label{d2}
\end{equation}
Representation $\left( \ref{d2}\right) $ shows a gravitational perturbation
decoupling. For a generic value of the angular momentum $L$, one gets

\begin{equation}
\left\{ 
\begin{array}{c}
-\triangle _lH\left( r\right) -\frac{4m}{r_{}^3}H\left( r\right)
=E_l^2H\left( r\right) \\ 
\\ 
-\triangle _lK\left( r\right) +\frac{2m}{r_{}^3}K\left( r\right)
=E_l^2K\left( r\right) \\ 
\\ 
-\triangle _lK\left( r\right) +\frac{2m}{r_{}^3}K\left( r\right)
=E_l^2K\left( r\right) .
\end{array}
\right.  \label{d3}
\end{equation}
The Laplacian restricted to $\Sigma $ can be written as

\begin{equation}
\triangle =\left( 1-\frac{2m}r\right) \frac{d^2}{dr^2}+\left( \frac{2r-3m}{%
r_{}^2}\right) \frac d{dr}-\frac{l\left( l+1\right) }{r_{}^2}.
\end{equation}
Defining reduced fields

\begin{equation}
H\left( r\right) =\frac{h\left( r\right) }r;\qquad K\left( r\right) =\frac{%
k\left( r\right) }r,
\end{equation}
and passing to the proper geodesic distance from the {\it throat} of the
bridge defined by 
\begin{equation}
dx=\pm \frac{dr}{\sqrt{1-\frac{2M}r}},  \label{p11}
\end{equation}
whose integrated form is

\begin{equation}
x=2m\left\{ \sqrt{\frac r{2m}}\sqrt{\frac r{2m}-1}+\ln \left( \sqrt{\frac r{%
2m}}+\sqrt{\frac r{2m}-1}\right) \right\} ,  \label{d2a}
\end{equation}
the system $\left( \ref{d3}\right) $ becomes

\begin{equation}
\left\{ 
\begin{array}{c}
-\frac{d^2}{dx^2}h\left( x\right) -V^{-}\left( x\right) h\left( x\right)
=E_l^2h\left( x\right)  \\ 
\\ 
-\frac{d^2}{dx^2}k\left( x\right) +V^{+}\left( x\right) k\left( x\right)
=E_l^2k\left( x\right)  \\ 
\\ 
-\frac{d^2}{dx^2}k\left( x\right) +V^{+}\left( x\right) k\left( x\right)
=E_l^2k\left( x\right) 
\end{array}
\right.   \label{d4}
\end{equation}
with 
\begin{equation}
V\left( x\right) =\frac{l\left( l+1\right) }{r^2\left( x\right) }\mp \frac{3m%
}{r\left( x\right) ^3}.
\end{equation}
This new variable represents the proper geodesic distance from the wormhole
throat such that 
\[
\text{when }r\longrightarrow \infty \text{, }x\simeq r\text{ , }V\left(
x\right) \longrightarrow 0
\]
\begin{equation}
\text{when }r\longrightarrow r_0\text{, }x\simeq 0,\text{ }V^{\mp }\left(
x\right) \longrightarrow \frac{l\left( l+1\right) }{r_0^2}\mp \frac{3m}{r_0^3%
}=const,
\end{equation}
where $r_0$ satisfies the condition $r_0>2m$. The solution of $\left( \ref
{d4}\right) $, in both cases (flat and curved one) is the spherical Bessel
function of the first kind 
\begin{equation}
j_0\left( px\right) =\sqrt{\frac 2\pi }\sin \left( px\right) 
\end{equation}
This choice is dictated by the requirement that 
\begin{equation}
h\left( x\right) ,k\left( x\right) \rightarrow 0\text{\qquad when\qquad }%
x\rightarrow 0\ \left( \text{alternatively }r\rightarrow 2m\right) .
\end{equation}
Then 
\begin{equation}
K\left( x,y\right) =\frac{j_0\left( px\right) j_0\left( py\right) }{2\lambda 
}\cdot \frac 1{4\pi }  \label{d5}
\end{equation}
Substituting $\left( \ref{d5}\right) $ in $\left( \ref{b6}\right) $ one gets
(after normalization in spin space and after a rescaling of the fields in
such a way as to absorb $l_p^2$) 
\begin{equation}
E\left( m,\lambda \right) =\frac V{2\pi ^2}\sum_{l=0}^\infty
\sum_{i=1}^2\int_0^\infty dpp^2\left[ \lambda _i\left( p\right) +\frac{%
E_i^2\left( p,m,l\right) }{\lambda _i\left( p\right) }\right]   \label{d6}
\end{equation}
where 
\begin{equation}
E_{1,2}^2\left( p,m,l\right) =p^2+\frac{l\left( l+1\right) }{r_0^2}\mp \frac{%
3m}{r_0^3},
\end{equation}
$\lambda _i\left( p\right) $ are variational parameters corresponding to the
eigenvalues for a (graviton) spin-two particle in an external field and $V$
is the volume of the system.

By minimizing $\left( \ref{d6}\right) $ with respect to $\lambda _i\left(
p\right) $ one obtains $\overline{\lambda }_i\left( p\right) =\left[
E_i^2\left( p,m,l\right) \right] ^{\frac 12}$ and 
\begin{equation}
E\left( m,\overline{\lambda }\right) =\frac V{2\pi ^2}\sum_{l=0}^\infty
\sum_{i=1}^2\int_0^\infty dp2\sqrt{E_i^2\left( p,m\right) }\text{ }
\end{equation}
with 
\[
\text{with }p^2+\frac{l\left( l+1\right) }{r_0^2}>\frac{3m}{r_0^3}.
\]
Thus, in presence of the curved background, we get 
\begin{equation}
E\left( m\right) =\frac V{2\pi ^2}\frac 12\sum_{l=0}^\infty \int_0^\infty
dpp^2\left( \sqrt{p^2+c_{-}^2}+\sqrt{p^2+c_{+}^2}\right)   \label{d7}
\end{equation}
where 
\[
c_{\mp }^2=\frac{l\left( l+1\right) }{r_0^2}\mp \frac{3m}{r_0^3},
\]
while when we refer to the flat space, in the spirit of the subtraction
procedure, we have $m=0$ and $c^2=$ $\frac{l\left( l+1\right) }{r_0^2}$.
Then 
\begin{equation}
E\left( 0\right) =\frac V{2\pi ^2}\frac 12\sum_{l=0}^\infty \int_0^\infty
dpp^2\left( 2\sqrt{p^2+c^2}\right)   \label{d8}
\end{equation}
Now, we are in position to compute the difference between $\left( \ref{d7}%
\right) $ and $\left( \ref{d8}\right) $. Since we are interested in the $UV$
limit, we have 
\[
\Delta E\left( m\right) =E\left( m\right) -E\left( 0\right) 
\]
\[
=\frac V{2\pi ^2}\frac 12\sum_{l=0}^\infty \int_0^\infty dpp^2\left[ \sqrt{%
p^2+c_{-}^2}+\sqrt{p^2+c_{+}^2}-2\sqrt{p^2+c^2}\right] 
\]
\begin{equation}
=\frac V{2\pi ^2}\frac 12\sum_{l=0}^\infty \int_0^\infty dpp^3\left[ \sqrt{%
1+\left( \frac{c_{-}}p\right) ^2}+\sqrt{1+\left( \frac{c_{+}}p\right) ^2}-2%
\sqrt{1+\left( \frac cp\right) ^2}\right] 
\end{equation}
and for $p^2>>c_{\mp }^2,c^2$, we obtain 
\[
\frac V{2\pi ^2}\frac 12\sum_{l=0}^\infty \int_0^\infty dpp^3\left[ 1+\frac 1%
2\left( \frac{c_{-}}p\right) ^2-\frac 18\left( \frac{c_{-}}p\right) ^4+1+%
\frac 12\left( \frac{c_{+}}p\right) ^2-\frac 18\left( \frac{c_{+}}p\right)
^4\right. 
\]
\begin{equation}
\left. -2-\left( \frac cp\right) ^2-\frac 14\left( \frac cp\right) ^4\right]
=-\frac V{2\pi ^2}\frac{c^4}8\int_0^\infty \frac{dp}p.
\end{equation}
We will use a cut-off $\Lambda $ to keep under control the $UV$ divergence 
\begin{equation}
\int_0^\infty \frac{dp}p\sim \int_0^{\frac \Lambda c}\frac{dx}x\sim \ln
\left( \frac \Lambda c\right) ,
\end{equation}
where $\Lambda \leq m_p.$ Thus $\Delta E\left( m\right) $ for high momenta
becomes 
\begin{equation}
\Delta E\left( m\right) \sim -\frac V{2\pi ^2}\frac{c^4}{16}\ln \left( \frac{%
\Lambda ^2}{c^2}\right) =-\frac V{2\pi ^2}\left( \frac{3m}{r_0^3}\right) ^2%
\frac 1{16}\ln \left( \frac{r_0^3\Lambda ^2}{3m}\right) .  \label{d10}
\end{equation}
{\bf Remark} It is known that at one-loop level Gravity is renormalizable
only in flat space. In a dimensional regularization scheme its contribution
to the action is, on shell, proportional to the Euler character of the
manifold that is nonzero for the Schwarzschild instanton. Although in our
approach we are working with sections of the original manifold to deal with
these divergences one must introduce a regulator that indeed appears in the
contribution of energy density, which we presume should be taken of order
the Planck mass $m_P\sim \sqrt{G^{-1}}$ in the same spirit of Ref.\cite{GPY}%
. This agrees with Wheeler's point of view concerning the gravitational
vacuum fluctuations, where a natural cut-off is introduced: the Planck
length $l_p\sim \sqrt{G}$\cite{Wheeler}. At this point we can compute the
total energy, namely the classical contribution plus the quantum correction
up to second order. Recalling the expression for quasilocal energy 
\begin{equation}
E_{{\rm quasilocal}}=\frac 1\kappa \int_{S_{+}}^{}d^2x\sqrt{\sigma }\left(
k-k^0\right) -\frac 1\kappa \int_{S_{-}}d^2x\sqrt{\sigma }\left(
k-k^0\right) ,  \label{c8}
\end{equation}
and by using the expression of the trace 
\begin{equation}
k=-\frac 1{\sqrt{h}}\left( \sqrt{h}n^\mu \right) _{,\mu },
\end{equation}
we obtain at either boundary that 
\begin{equation}
k=\frac{-2r,_y}r,
\end{equation}
where we have assumed that the function $r,_y$ is positive for $S_{+}$ and
negative for $S_{-}$. The trace associated with the subtraction term is
taken to be $k^0=-2/r$ for $B_{+}$ and $k^0=2/r$ for $B_{-}$. Then the
quasilocal energy with subtraction terms included is 
\begin{equation}
E_{{\rm quasilocal}}=E_{+}-E_{-}=\left( r\left[ 1-\left| r,_y\right| \right]
\right) _{y=y_{+}}-\left( r\left[ 1-\left| r,_y\right| \right] \right)
_{y=y_{-}}.
\end{equation}
Note that the total quasilocal energy is zero for boundary conditions
symmetric with respect to the bifurcation surface $S_0$. The energy $%
E_{+}\left( E_{-}\right) $ tends to the ${\cal ADM}$ mass $M$\cite{ADM}
whenever the boundary $B_{+}\left( B_{-}\right) $ tends to right (left)
spatial infinity. As an illustration, consider the case when the boundary $%
B_{+}$ is located at right-hand infinity $\left( y_{+}=+\infty \right) $ and
the boundary $B_{-}$ is located at $y_{-}$. The total energy for the stable
modes is 
\begin{equation}
M-r\left[ 1-\left( 1-\frac{2MG}r\right) ^{\frac 12}\right] -\frac V{2\pi ^2}%
\left( \frac{3MG}{r_0^3}\right) ^2\frac 1{16}\ln \left( \frac{r_0^3\Lambda
_{}^2}{3MG}\right) ,  \label{c9}
\end{equation}
while the total square energy for the unstable modes is 
\[
E^2=-\frac{a^2}{8\left( MG\right) ^2}
\]
with $a^2=0,242$ and will be computed in the appendix \ref{p7}. Then the one
loop total energy is 
\begin{equation}
M-r\left[ 1-\left( 1-\frac{2MG}r\right) ^{\frac 12}\right] -\frac V{2\pi ^2}%
\left( \frac{3MG}{r_0^3}\right) ^2\frac 1{16}\ln \left( \frac{r_0^3\Lambda
_{}^2}{3MG}\right) +a\frac i{MG}.  \label{c3}
\end{equation}
If we consider only one of the wedges, we obtain the correction relative to
the positive (negative) ${\cal ADM}$ mass $M$ , depending on the location of
the wedge. One can observe that 
\begin{equation}
\Delta E\left( M\right) \rightarrow \infty \text{ when }M\rightarrow 0\text{%
, for }r_0=2GM
\end{equation}
and 
\begin{equation}
\Delta E\left( M\right) \rightarrow 0\text{ when }M\rightarrow 0\text{, for }%
r_0\neq 2GM.
\end{equation}
Note that this singular behaviour is independent of boundary conditions
since it is related to the volume term.

\section{Summary and Conclusions}

\label{p4}

We started from the problem of defining (semiclassical) quantum corrections
to the quasilocal energy. By means of a variational approach with Gaussian
wave functionals, an attempt to calculate such a correction was made. By
construction, we used the subtraction procedure given in Refs.\cite
{BrownYork,FroMar,HawHor} to avoid divergences coming from boundaries.
Despite the constraint equations, this calculation is based on an extension
of the subtraction procedure involving volume terms in the semiclassical
regime. Excitations coming from boundary terms have been neglected to avoid
the unphysical situation of having contributions deriving from infinity. In
this context, the extended subtraction procedure corresponds to the
difference between zero point energies calculated in an asymptotically flat
background referring to a flat background. This procedure eliminates the UV
divergence of the free gravitons, leaving the contribution of the curved
background related to an {\it imposed by hand} UV cut-off. Note that the
subtraction procedure at one loop has the correct ingredients to be related
to the Casimir energy. This seems to give some information about the vacuum
behaviour. Indeed, if we look at symmetric boundary conditions with respect
to the bifurcation surface $S_0$, then eq.$\left( \ref{c3}\right) $ can be
interpreted as an energy gap measuring the probability of creating a black
hole pair by a topological fluctuation; it also indicates that flat
spacetime is unstable with respect to pair creation. In the introduction we
mentioned the impossibility of generating single black holes by quantum
fluctuation in a flat space having zero temperature, because the energy
would not be conserved\cite{GPY,Witten}. However, in the neutral black hole
pair scenario\cite{Remo}, where each component resides in a different
universe, the energy is conserved provided that the boundaries are symmetric
with respect to the bifurcation $S_0$. Note that this calculation can be
related to the quantity 
\begin{equation}
\Gamma _{{\rm 1-hole}}=\frac{P_{{\rm 1-hole}}}{P_{{\rm flat}}}\simeq \frac{%
P_{{\rm BlackHolePair}}}{P_{{\rm flat}}}.  \label{s1}
\end{equation}
To generalize a little more, suppose to enlarge this process from one pair
to a large but fixed number of such pairs, say $n$. What we obtain is a
multiply connected spacetime with $n$ holes inside the manifold, each of
them acting as a single bifurcation surface with the sole condition of
having symmetry with respect to the bifurcation surface even at finite
distance. Let us suppose the interaction between the holes can be neglected,
i.e., let us suppose that the total energy contribution is realized with a
coherent summation process. This is equivalent to saying that the wave
functional support (here, the semiclassical WDW functional) has a finite
size depending only on the number of the holes inside the spacetime. It is
clear that the number of such holes cannot be arbitrary, but is to be
related with a minimum size of Planck's order. Thus, assuming a coherency
property of the wave functional and therefore of the $N$-holes spacetime, eq.%
$\left( \ref{s1}\right) $ has to be generalized to 
\begin{equation}
\Gamma _{{\rm N-holes}}=\frac{P_{{\rm N-holes}}}{P_{{\rm flat}}}\simeq \frac{%
P_{{\rm foam}}}{P_{{\rm flat}}}.  \label{s2}
\end{equation}
Recall that a hole, here, has to be understood as a wormhole. Then, what eq.$%
\left( \ref{s2}\right) $ suggests is the possibility of generating a {\it %
foamy} spacetime, with wormholes as building blocks.

\appendix

\section{Conventions and Scalar Curvature Expansion}

\label{p5}

\begin{itemize}
\item  Riemann tensor 
\begin{equation}
R_{ijm}^l=\Gamma _{mi,j}^l-\Gamma _{ji,m}^l+\Gamma _{ja}^l\Gamma
_{mi}^a-\Gamma _{ma}^l\Gamma _{ji}^a.
\end{equation}

\item  Ricci tensor 
\begin{equation}
R_{im}=R_{ilm}^l.
\end{equation}

\item  Scalar curvature 
\begin{equation}
R=g_{}^{lj}R_{lj}^{}.
\end{equation}

\item  In three dimensions, the Weyl tensor vanishes, then the Riemann
tensor is completely determined by Ricci tensor 
\begin{equation}
R_{lijm}^{}=g_{lj}R_{im}^{}-g_{lm}R_{ij}^{}-g_{ij}R_{lm}^{}+g_{im}R_{lj}^{}.
\end{equation}

\item  Second order scalar curvature
\end{itemize}

\begin{equation}
\int d_{{}}^{3}x\left[ -\frac{1}{4}h\triangle h+\frac{1}{4}h^{li}\triangle
h_{li}-\frac{1}{2}h^{ij}\nabla _{l}\nabla _{i}h_{j}^{l}+\frac{1}{2}h\nabla
_{l}\nabla _{i}h_{{}}^{li}-\frac{1}{2}h^{ij}R_{ia}h_{j}^{a}+\frac{1}{2}%
hR_{ij}h_{{}}^{ij}\right] .  \label{app4}
\end{equation}

\section{The Kinetic Term}

\label{p6}

The Schr\"{o}dinger picture representation of the kinetic term is 
\begin{equation}
G_{ijkl}\pi ^{ij}\pi ^{kl}=G_{ijkl}\left( -\frac{\delta ^2}{\delta
h_{ij}\left( x\right) \delta h_{kl}\left( x\right) }\right) .
\end{equation}
We have to apply this quantity to the gaussian wave functional $\left| \Psi
\right\rangle $. This means that 
\[
\pi ^{ij}\left( x\right) \pi ^{kl}\left( x\right) \left| \Psi \right\rangle
=-\frac{\delta ^2\Psi \left[ h\right] }{\delta h_{ij}\left( x\right) \delta
h_{kl}\left( x\right) } 
\]
\[
=\frac 12K^{-1\left( kl\right) \left( ij\right) }\left( x,x\right) \left( 
\sqrt{g\left( x\right) }\right) ^2\Psi \left[ h\right] 
\]
\[
-\frac 14\int d^3y^{\prime }d^3y^{\prime \prime }\left( \sqrt{g\left(
x\right) }\right) ^2\sqrt{g\left( y^{\prime }\right) }\sqrt{g\left(
y^{\prime \prime }\right) }K^{-1\left( kl\right) \left( k^{\prime }l^{\prime
}\right) }\left( x,y^{\prime }\right) h_{k^{\prime }l^{\prime }}\left(
y^{\prime }\right) 
\]
\begin{equation}
\cdot K^{-1\left( ij\right) \left( k^{\prime \prime }l^{\prime \prime
}\right) }\left( x,y^{\prime \prime }\right) h_{k^{\prime \prime }l^{\prime
\prime }}\left( y^{\prime \prime }\right) \Psi \left[ h\right] .  \label{ab1}
\end{equation}
By functional integration 
\begin{equation}
\left\langle \Psi \left| h_{k^{\prime }l^{\prime }}\left( y^{\prime }\right)
h_{k^{\prime \prime }l^{\prime \prime }}\left( y^{\prime \prime }\right)
\right| \Psi \right\rangle =K_{\left( k^{\prime }l^{\prime }\right) \left(
k^{\prime \prime }l^{\prime \prime }\right) }\left( y^{\prime },y^{\prime
\prime }\right) \left\langle \Psi |\Psi \right\rangle .
\end{equation}
Then 
\[
\left\langle \Psi \left| \pi ^{ij}\left( x\right) \pi ^{kl}\left( x\right)
\right| \Psi \right\rangle 
\]
becomes 
\[
\frac 12K^{-1\left( kl\right) \left( ij\right) }\left( x,x\right) \left( 
\sqrt{g\left( x\right) }\right) ^2 
\]
\[
-\frac 14\int d^3y^{\prime }d^3y^{\prime \prime }\left( \sqrt{g\left(
x\right) }\right) ^2\sqrt{g\left( y^{\prime }\right) }\sqrt{g\left(
y^{\prime \prime }\right) }K^{-1\left( kl\right) \left( k^{\prime }l^{\prime
}\right) }\left( x,y^{\prime }\right) K^{-1\left( ij\right) \left( k^{\prime
\prime }l^{\prime \prime }\right) }\left( x,y^{\prime \prime }\right) 
\]
\[
K_{\left( k^{\prime }l^{\prime }\right) \left( k^{\prime \prime }l^{\prime
\prime }\right) }\left( y^{\prime },y^{\prime \prime }\right) \left\langle
\Psi |\Psi \right\rangle 
\]
\begin{equation}
=\frac 14K^{-1\left( kl\right) \left( ij\right) }\left( x,x\right) \left( 
\sqrt{g\left( x\right) }\right) ^2\left\langle \Psi |\Psi \right\rangle .
\end{equation}
Then the expectation value of the kinetic term, with the Planck length
reinserted, is 
\begin{equation}
\left\langle T\right\rangle =\frac 1{4l_p^2}\int d^3x\sqrt{g}\left(
G_{ijkl}K^{-1\left( kl\right) \left( ij\right) }\left( x,x\right) \right) ,
\label{ab2}
\end{equation}

\section{Searching for negative modes}

\label{p7}

In this paragraph we look for negative modes of the eigenvalue equation $%
\left( \ref{d1}\right) $. For this purpose we restrict the analysis to the S
wave. Indeed, in this state the centrifugal term is absent and this gives
the function $V\left( x\right) $ a potential well form, which is different
when $l\geq 1$, where $l$ is the angular momentum. However, this potential
form is valid only for the $H$ component, i.e. 
\[
-\triangle H\left( r\right) -\frac{4m}{r^3}H\left( r\right) =-E^2H\left(
r\right) ,
\]
with $E^2>0$. Passing to the reduced field $H\left( r\right) =\frac{h\left(
r\right) }r$, we obtain 
\[
-\frac d{dr}\left( \sqrt{1-\frac{2m}r}\frac{dh}{dr}\right) +\left( \frac{-3m%
}{r^3}+E^2\right) \frac h{\sqrt{1-\frac{2m}r}}=0.
\]
Making the substitution $\left( \ref{d2a}\right) ,$ the equation becomes 
\[
-\frac{dx}{dr}\frac d{dx}\left( \sqrt{1-\frac{2m}r}\frac{dh}{dx}\frac{dx}{dr}%
\right) +\left( -\frac{3m}{r^3}+E^2\right) \frac h{\sqrt{1-\frac{2m}r}}
\]
\[
=-\frac d{dx}\left( \frac{dh}{dx}\right) +\left( -\frac{3m}{r^3}+E^2\right)
h=0,
\]
where $\frac{dx}{dr}=\frac 1{\sqrt{1-\frac{2m}r}}.$ Near the horizon $%
x\simeq 4m\sqrt{\frac r{2m}-1}$or $r\simeq 2m\left( 1+\left( \frac x{4m}%
\right) ^2\right) .$ Define $\rho =\frac r{2m}\Longrightarrow \rho =1+\left( 
\frac y2\right) ^2$ with $y=\frac x{2m}.$ Then we get 
\[
-\frac d{dy}\left( \frac{dh}{dy}\right) +\left( -\frac{3m}{\left( 2m\right)
^3\rho ^3\left( y\right) }+E^2\right) h\rightarrow -\frac{d^2h}{dy^2}+\left(
-\frac 3{2\left( 1+\left( \frac y2\right) ^2\right) ^3}+\lambda \right) h=0,
\]
where $\lambda =\left( 2m\right) ^2E^2.$ Expanding the potential around $y=0$%
, one gets 
\[
-\frac{d^2h}{dy^2}+\left( -\frac 32\left( 1-\frac 34y^2\right) +\lambda
\right) h=0.
\]
We see that this is a quantum harmonic oscillator equation whose spectrum is
well known, that is $E_n=\hbar \omega \left( n+\frac 12\right) .$ However,
we have to remark that the original spectrum must be contained in the
interval , where $-\frac 32$ represents the bottom of the approximated
potential. Then $E_n=\frac 32+\lambda _n=\sqrt{\frac 38}\left( n+\frac 12%
\right) $, where $\omega =\sqrt{\frac 38}$ and $\hbar =1$ in natural units.
Thus 
\[
\lambda _n=-\frac 32+\frac 32\frac{\sqrt{2}}2\left( n+\frac 12\right) .
\]
We see that 
\[
\lambda _0=-0.975\text{ and }\lambda _1=-\frac 32\left( -0.06\right)
\Longrightarrow \lambda _1\notin \left( -\frac 32,0\right) .
\]
There is only {\bf one eigenvalue. }The same problem can be approached with
the Rayleigh-Ritz method along a numerical integration and the result is $%
\lambda _0=-1.094$.


\begin{references}
\bibitem{Yau}  R. Schoen, S.T. Yau, Commun. Math. Phys. {\bf 65}, 45 (1979);
Commun. Math. Phys. {\bf 79}, 231 (1981).

\bibitem{GPY}  D.J. Gross, M.J. Perry and L.G. Yaffe, Phys. Rev. D {\bf 25},
(1982) 330.

\bibitem{Remo}  R. Garattini, {\it Spacetime Foam, Casimir Energy and Black
Hole Pair Creation, }Mod. Phys. Lett. A {\bf 13}, 159 (1998), Report
gr-qc/98010045.

\bibitem{BrownYork}  J.D. Brown and J.W. York, Phys. Rev. D {\bf 47}, 1407
(1993).

\bibitem{FroMar}  V.P. Frolov and E.A. Martinez, {\it Action and Hamiltonian
for Eternal Black Holes, }Class.Quant.Grav.13 :481-496,1996, gr-qc/9411001.

\bibitem{HawHor}  S. W. Hawking and G. T. Horowitz, {\it The Gravitational
Hamiltonian, Action, Entropy and Surface Terms}, Class. Quant. Grav. {\bf 13 
}1487, (1996), Report gr-qc/9501014.

\bibitem{BerEbi}  M. Berger and D. Ebin, J. Diff. Geom. {\bf 3}, 379 (1969).

\bibitem{York}  J. W. York Jr., J. Math. Phys., {\bf 14}, 4 (1973); Ann.
Inst. Henri Poincar\'{e} {\bf A21}, 319{\bf \ }(1974).

\bibitem{HalHaw}  J.J. Halliwell and S.W. Hawking, Phys. Rev. {\bf 31}, 1777
(1985).

\bibitem{Halliwell}  J.J. Halliwell, ``Introductory Lectures on Quantum
Cosmology''. In {\it Jerusalem Winter School for Theoretical Physics:
Quantum Cosmology and Baby Universes Vol. 7}. S.Coleman, J.B. Hartle, T.
Piran and S. Weinberg, eds. World Scientific, 159-243.

\bibitem{Brotz}  T. Brotz, {\it Quantization of Black Holes in the
Wheeler-DeWitt Approach}, Report gr-qc/9708066.

\bibitem{ADM}  R. Arnowitt, S. Deser, and C. W. Misner, in {\it Gravitation:
An Introduction to Current Research,} edited by L. Witten (John Wiley \&
Sons, Inc., New York, 1962); B. S. DeWitt, Phys. Rev. {\bf 160}, 1113 (1967).

\bibitem{CJT}  A. K. Kerman and D. Vautherin, Ann. Phys., {\bf 192}, 408
(1989); J. M. Cornwall, R. Jackiw and E. Tomboulis, Phys. Rev. D {\bf 8},
2428 (1974); R. Jackiw in {\it S\'{e}minaire de Math\'{e}matiques
Sup\'{e}rieures, Montr\'{e}al, Qu\'{e}bec, Canada- June 1988 - Notes by P.
de Sousa Gerbert}; M. Consoli and G. Preparata, Phys. Lett. B, {\bf 154},
411 (1985).

\bibitem{Regge}  T. Regge and J. A. Wheeler, Phys. Rev. {\bf 108}, 1063
(1957).

\bibitem{Wheeler}  J. A. Wheeler, {\it Geometrodynamics} (Ac. Press, London,
1962).

\bibitem{Witten}  E. Witten, Nucl. Phys. B {\bf 195 (}1982) 481.

\bibitem{MTW}  C.W.Misner, K.S. Thorne and J.A. Wheeler, {\it Gravitation}
(Freeman, San Francisco, 1973) 842.
\end{references}
\end{document}